\documentclass[conference]{IEEEtran}
\IEEEoverridecommandlockouts
\usepackage{cite}
\usepackage{amsmath,amssymb,amsfonts}
\usepackage[ruled,linesnumbered]{algorithm2e}
\usepackage{graphicx}
\usepackage{textcomp}
\usepackage{xcolor}
\usepackage{threeparttable}
\usepackage{booktabs}
\usepackage{multirow}
\usepackage{bm}
\usepackage{threeparttable}
\usepackage{tikz}
\usetikzlibrary{shapes,arrows.meta,positioning,calc,fit,backgrounds}

\def\BibTeX{{\rm B\kern-.05em{\sc i\kern-.025em b}\kern-.08em
    T\kern-.1667em\lower.7ex\hbox{E}\kern-.125emX}}

\newcommand{\vect}[1]{\mathbf{#1}}

\begin{document}


\title{CSI RefineNet: A Soft Data-Aided Iterative Receiver for High-Mobility OFDM Systems}
\author{\IEEEauthorblockN{Yirui Luo$^*$, Yihang Xie$^*$, Yao Ge$^*$, Yong Liang Guan$^*$, and David~Gonz\'{a}lez~G.$^\dagger$}
\IEEEauthorblockA{
\textit{$^*$AUMOVIO-NTU Corporate Lab, Nanyang Technological University, Singapore } \\
\textit{$^\dagger$Wireless Communications Technologies Group, AUMOVIO Germany GmbH} \\
yirui001@e.ntu.edu.sg, $\{$yihang.xie; yao.ge; eylguan$\}$@ntu.edu.sg, david.gonzalez.g@ieee.org}
}

\maketitle

\begin{abstract}
    In high-mobility orthogonal frequency division multiplexing (OFDM) systems, rapid channel variation can make the channel state information (CSI) estimated from pilots inaccurate for data subcarriers, leading to a mismatch with their effective channel. To address this issue, this paper proposes a CSI RefineNet receiver, where the CSI is iteratively refined using soft symbol decisions in a data-aided manner.
    Specifically, a pilot-driven initialization module is first employed to obtain a coarse CSI estimation and the corresponding symbol posterior probabilities. Based on these posteriors, soft data-aided channel observations are constructed over all subcarriers and fused with the initial CSI to refine the channel estimation. The refined CSI is subsequently fed back to the equalization and detection modules, thereby forming an iterative receiver structure. To improve training stability and fully exploit the refinement capability, a two-stage training strategy is also developed. Simulation results demonstrate that the proposed CSI RefineNet receiver achieves superior BER performance and strong robustness under different velocities, modulation orders, and pilot spacing configurations in high-mobility OFDM systems.
\end{abstract}


\section{Introduction}
\label{sec:intro}

Orthogonal frequency-division multiplexing (OFDM) is widely adopted in broadband wireless systems because of the high spectral efficiency and simple frequency-domain equalization. However, in high-mobility scenarios, severe Doppler spread causes the wireless channel to vary significantly even within one OFDM symbol duration. As a result, the orthogonality among subcarriers is destroyed, leading to inter-carrier interference (ICI) and degrading the reliability of conventional pilot-aided channel estimation methods \cite{jeon1999equalization, mostofi2005ici,MYIOTJ10963873}. These issues are particularly critical in high-speed railway and vehicle-to-everything (V2X) links, as in several 6G scenarios \cite{3GPP}, accurate channel state information (CSI) must be continuously maintained under rapidly time-varying propagation conditions.

Classical estimators, such as least squares (LS) and minimum mean square error (MMSE)  remain attractive because of their simplicity and analytical structure \cite{edfors1998ofdm}. However, their performance is limited when pilot density is low and the channel exhibits strong time selectivity within the OFDM symbol \cite{tang2007pilot}. 
Recent work has shown that deep learning can improve OFDM channel estimation and detection by exploiting structure in the time-frequency observations. For example, \cite{ye2018power} showed the potential of neural networks for improving channel estimation and signal detection. Subsequently, \cite{soltani2019deep} proposed the ChannelNet framework, which reformulated channel estimation as an image super-resolution task and provided a novel deep-learning-based approach for CSI reconstruction.
Authors in \cite{xie2024commtransformer} further proposed the transformer-based receiver to improve the ability to capture positional information across subcarriers. 
Motivated by advances in learning-based physical-layer design, the authors in \cite{honkala2021deeprx} proposed DeepRx, which demonstrates strong detection performance in matched settings.
Later works further investigated learning-based receivers in more challenging scenarios, such as those involving nonlinear distortion \cite{xie2022nonlinear} and severe doubly selective fading channels \cite{liu2024bemann} \cite{OTFS_10516684}.

Although the aforementioned neural receivers improve robustness to some extent, they still rely on a single-pass reception framework. Pilot-based neural estimators use only the pilot tones, and end-to-end receivers usually lack an explicit mechanism to refine the CSI after tentative symbol decisions become available. Consequently, most existing robust receiver designs still perform a one-pass mapping from the received observation directly to channel or symbol estimates. Moreover, fully black-box architectures may generalize less reliably when channel statistics change \cite{cammerer2020trainable}. By contrast, classical iterative receivers have long benefited from decision-directed processing, where tentative decisions are fed back to improve intermediate estimates \cite{wang1999iterative, tuchler2002mmse, sandell1998iterative}. Related model-driven and unfolded detectors further highlight the value of combining domain structure with learning \cite{samuel2019learning, khani2020adaptive, he2020model}, while iterative Bayesian detectors such as generalized approximate message passing (GAMP) \cite{GAMP_8746319} can improve symbol inference but still rely on a fixed channel estimate from a separate estimator.
We believe the key is not just relying on one single-shot detection, but using soft decisions to sharpen CSI before each new detection round.

\begin{figure*}
    \centering
    \includegraphics[width=0.85\linewidth]{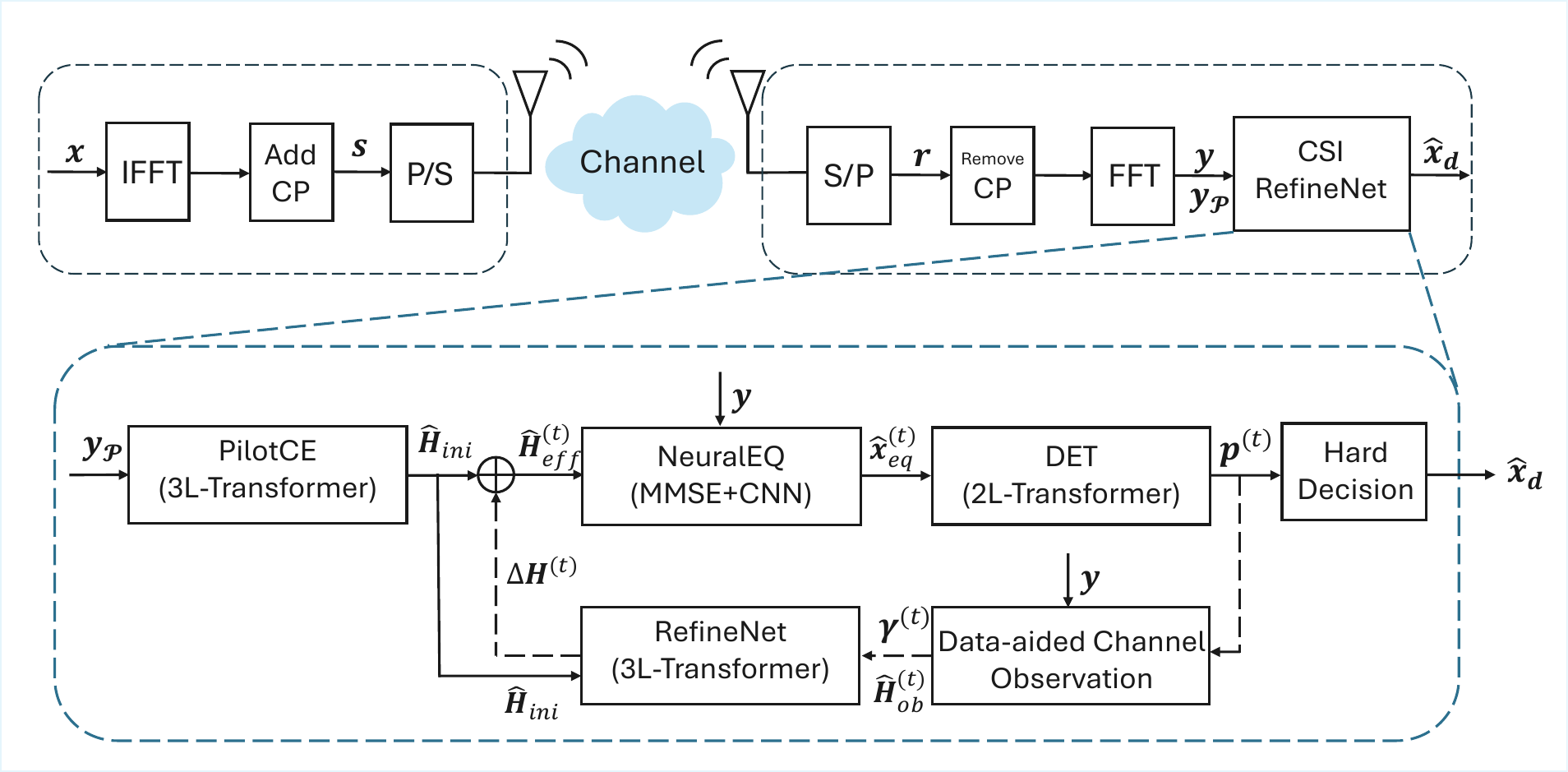}
    \caption{Block diagram of the proposed OFDM transceivers.}
    \label{Receiver_Arc}
\end{figure*}

Such performance gap motivates us to develop a receiver that bridges monolithic end-to-end learning and fully model-based iterative processing. In this paper, we propose a CSI RefineNet receiver for high-Doppler OFDM scenarios, which first obtains a coarse CSI estimate from pilots, and then exploits detection outputs to form data-aided channel observations for iterative CSI refinement through a dedicated refinement network. The refined CSI is fed back to the equalization and detection modules, forming a closed-loop receiver. Unlike existing single-pass learned receivers, CSI RefineNet treats the intermediate CSI as a quantity that can be progressively improved once tentative decisions become reliable. 


\section{System Model}
\label{sec:system}

We consider a single-input single-output (SISO) OFDM system with carrier frequency $f_c$ and bandwidth $B=N\Delta f$, where $N$ denotes the number of subcarriers and $\Delta f$ is the subcarrier spacing. Pilot symbols are inserted at positions $\mathcal{P} = \{0, P_s, 2P_s, \ldots\}$ with spacing $P_s$. The block diagram of our system transceiver is illustrated in Fig.~\ref{Receiver_Arc}.
\subsection{OFDM Transmission}
Let $\mathbf{x} = \left[x[0], x[1], \ldots, x[N-1]\right]^T \in \mathbb{C}^{N\times 1}$ denote the transmitted symbols in the frequency domain, where $x[k] \in \mathcal{C}$ is drawn from an $M$-ary constellation set $\mathcal{C}$. Then, the time domain signal is obtained by an $N$-point inverse fast Fourier transform (IFFT) with the expression of
\begin{equation}
    s[n] = \frac{1}{\sqrt{N}} \sum_{k=0}^{N-1} x[k] e^{j2\pi kn/N}, 
    \label{eq:ifft}
\end{equation}
with $n = 0, \ldots, N-1$. 
To avoid inter-symbol interference (ISI), a cyclic prefix (CP) of length $N_{cp}$ is appended by prepending the last $N_{cp}$ samples of $\mathbf{s}$ to its beginning, where 
$\mathbf{s} = \left[s[0], s[1], \ldots, s[N-1]\right]^T \in \mathbb{C}^{N\times 1}$ and $N_{cp}$ is no smaller than the channel delay spread. Then, the transmitted time domain OFDM signal with CP $\mathbf{s} \in \mathbb{C}^{\left(N+N_{cp}\right)\times 1}$ can be expressed as
\begin{equation}
    s\left[n\right] =
\begin{cases}
s\left[n + N\right], & n = -N_{cp}, \dots, -1, \\
s\left[n\right], & n = 0, 1, \dots, N-1.
\end{cases}
\end{equation}
\subsection{Received Signal}

Propagation takes place over a time-varying multipath channel with $L$ taps. Let $\alpha_l$, $\tau_l$ and $\nu_l$ denote the complex channel gain, {delay in samples}, and Doppler shift of the $l$-th tap, respectively. 
The received sample at time $n$ is then written as
\begin{equation}
r\left[n\right]=\sum_{l=0}^{L-1}\alpha_l e^{j2\pi \nu_l n T_s}\, s\left[n-\tau_l\right]+w_t\left[n\right],
\end{equation}
where $n = 0, \ldots ,\left( {N + {N_{cp}}} \right) - 1$, $T_s=1/B$ is the sampling interval and $w_t[n] \sim \mathcal{CN}(0,\sigma_w^2)$ is additive white Gaussian noise.

After CP removal and FFT, the received symbol on subcarrier $k$ can be expressed as
\begin{equation}
y\left[k\right]=\sum_{m=0}^{N-1} H\left[k,m\right]x\left[m\right]+w[k],
\end{equation}
where $k=0,...,N-1$, $w[k]$ denotes the frequency-domain noise term, and $H[k,m]$ is the effective channel response from the $m$-th transmitted subcarrier to the $k$-th received subcarrier, given by
\begin{equation}
\begin{split}
    H\left[k,m\right]=&\frac{1}{N}\sum_{l=0}^{L-1}\alpha_l e^{-j\frac{2\pi m\tau_l}{N}}
    e^{j2\pi \nu_l N_{cp}T_s}\\
    &\times \sum_{n=0}^{N-1}e^{j2\pi \nu_l nT_s}
    e^{-j\frac{2\pi (k-m)n}{N}}.
\end{split}
\end{equation}
Accordingly, the received signal on the \(k\)-th subcarrier in the frequency domain can be decomposed into the desired term, the inter-carrier interference (ICI), and noise, which is given by

\begin{equation}
    y[k] = H[k,k]x[k] + \underbrace{\sum_{m \neq k} H[k,m] x[m]}_{\text{ICI}} + w[k],
    \label{y_freq}
\end{equation}
for $k=0,1,...,N-1$.
When the channel is approximately static over the OFDM symbol, the ICI term in \eqref{y_freq} becomes negligible and $\mathbf{H}$ reduces to the familiar diagonal model. However, in the considered time-varying channel with Doppler shifts, the ICI term is generally non-negligible. To retain a diagonal-form representation, we define an effective diagonal matrix $\mathbf{H}_{eff}\in \mathbb{C}^{N\times N}$, whose $k$-th diagonal element is given by
\begin{equation}
    H_{eff}[k,k]=H[k,k]+\frac{\sum_{m\ne k}H[k,m]x[m]}{x[k]}.
\end{equation}
Then, the received signal in \eqref{y_freq} can be rewritten as
\begin{equation}
    y[k]=H_{eff}[k,k]x[k]+w[k].
\end{equation}



\section{Proposed CSI RefineNet Receiver}
\label{sec:method}

\subsection{Architecture Overview}
As illustrated in Fig.~\ref{Receiver_Arc}, the proposed CSI-RefineNet receiver consists of four main components: PilotCE, EQ, DET, and RefineNet. 
Specifically, PilotCE operates only on $\vect{y}_{\mathcal{P}} \in \mathbb{C}^{|\mathcal{P}|\times 1}$ to recover coarse CSI with the expression of 
\begin{equation}
    \hat{\mathbf{H}}_{ini} = f_{\mathrm{CE}}(\mathbf{y}_{\mathcal{P}}; \theta_{\mathrm{CE}}), \label{H_ini}
\end{equation}
where $f_{\mathrm{CE}}(\cdot; \theta_{\mathrm{CE}})$ denotes the PilotCE network and $\theta_{\mathrm{CE}}$ denotes its trainable parameters. The full OFDM observation is represented as $\vect{y} \in \mathbb{C}^{N \times 1}$.

Then, the CSI update is performed in an iterative manner. At each iteration, NeuralEQ and DET generate tentative symbol posteriors based on the current CSI estimate, which are then used to form data-aided channel observations. These observations are fed into RefineNet to refine the residual mismatch between pilot and data components. The refined CSI is subsequently reused by the equalizer–detector pair, and this process is repeated iteratively.

In the $t$-th iteration, the output of the NeuralEQ and DET can be respectively expressed as 
\begin{equation}
    \mathbf{\hat x}_{eq}^{(t)} = f_{\mathrm{EQ}}\left(\mathbf{y}, \hat{\mathbf{H}}_{eff}^{(t)}; \theta_{\mathrm{EQ}}\right), \label{x_eq}
\end{equation}
\begin{equation}
    \mathbf{p}^{(t)} = f_{\mathrm{DET}}\left(\mathbf{\hat x}_{eq}^{(t)}; \theta_{\mathrm{DET}}\right),\label{p_det}
\end{equation}
where $f_{\mathrm{EQ}}(\cdot; \theta_{\mathrm{EQ}})$ and $f_{\mathrm{DET}}(\cdot; \theta_{\mathrm{DET}})$ denote the NeuralEQ module and DET module, respectively. Notations $\theta_{\mathrm{EQ}}$ and $\theta_{\mathrm{DET}}$ denote their corresponding trainable parameters.
Here, $\hat{\mathbf{H}}_{eff}^{(1)}=\hat{\mathbf{H}}_{ini}$ for the first iteration and $\mathbf{p}_k^{(t)} \in \mathbb{R}^{1\times M}$ is the $k$-th row of $\mathbf{p}^{(t)} \in \mathbb{C}^{N\times M}$, which denotes the posterior probability vector for subcarrier $k$.

In the data-aided channel observation procedure, the posterior probabilities are first mapped to soft symbol estimations as
\begin{equation}
    \tilde{x}[k] = \sum_{m=1}^{M} p^{(t)}[k,m] c[m], \quad k \notin \mathcal{P},
    \label{soft_sym} 
\end{equation}
where $p^{(t)}[k,m]$ is the $\left(k,m\right)$-th element of $\mathbf{p}^{(t)}$ and $c[m] \in \mathcal{C}$ denotes the $m$-th constellation point. For pilot positions, the transmitted symbols are known in a prior, and thus we directly set $\tilde{x}[k]= x[k]$, $k\in \mathcal{P}$.
Based on the obtained soft symbols, a data-aided channel observation matrix $\mathbf{\hat H}_{ob}\in \mathbb{C}^{N\times N}$ is constructed over all subcarriers, which is diagonal with its $k$-th diagonal entry given by
\begin{equation}
    \hat{H}_{{ob}}[k,k] = \frac{y[k]}{\tilde{x}[k]}, \label{H_ob}
\end{equation}
for $k=0,...,N-1$.
Compared with hard re-modulation, this soft construction preserves the uncertainty associated with unreliable symbol decisions, thereby mitigating the risk of error propagation in the subsequent refinement stage. A confidence score is also assigned to each subcarrier as
\begin{equation}
    \gamma[k] =
    \begin{cases}
        1, & k \in \mathcal{P}, \\
        \max_m p^{(t)}[k,m], & k \notin \mathcal{P}.
    \end{cases}
    \label{conf}
\end{equation}

The initial estimate, the data-aided observation, and the confidence vector are concatenated to form a multi-channel input tensor, where the real and imaginary components are treated separately, i.e.,
\begin{equation}
     \vect{Z}^{(t)} = \left[\mathrm{Re}\left(\hat{\vect{H}}_{ini}\right), \mathrm{Im}\left(\hat{\vect{H}}_{ini}\right),\mathrm{Re}\left(\hat{\vect{H}}^{(t)}_{ob}\right), \mathrm{Im}\left(\hat{\vect{H}}^{(t)}_{ob}\right), \bm{\gamma}\right].\label{Z_T}
\end{equation}

The resulting tensor is then fed into RefineNet, which produces a refinement term with the expression of 
\begin{equation}
    \Delta \mathbf{H} = f_{\mathrm{Ref}}(\vect{Z}; \theta_{\mathrm{Ref}}),\label{Delta_H}
\end{equation}
where $f_{\mathrm{Ref}}(\cdot; \theta_{\mathrm{Ref}})$ denotes the RefineNet module and $\theta_{\mathrm{Ref}}$ denotes its trainable parameters. The effective CSI is subsequently updated in a residual manner as
\begin{equation}
    \hat{\vect{H}}_{eff}^{(t)} = \hat{\vect{H}}_{ini} + \Delta \hat{\vect{H}}^{(t)}.
\end{equation}

\SetAlgoVlined
\begin{algorithm}[t]
\caption{Online Inference Procedure of Proposed CSI RefineNet Receiver}
\label{alg_refine}

\KwIn{Received signal $\vect{y}$, pilot observations $\vect{y}_{\mathcal{P}}$, refinement steps $T$}

\textbf{Initialization}:
$t=0$, ${\Delta \mathbf{H} ^{\left( 0 \right)}} = \mathbf{0}$\;
Obtain the pilot-driven coarse CSI matrix $\hat{\vect{H}}_{ini}$ according to \eqref{H_ini}\;

\Repeat{$t = T$}{
    $t\gets t+1$\;
    $\mathbf{H}_{eff}^{(t)}=\hat{\vect{H}}_{ini}+\Delta \mathbf{H}^{(t-1)}$\;
    Obtain the output of the NeuralEQ ${\mathbf{\hat x}}_{eq}^{(t)}$ via \eqref{x_eq}\;
    Get posterior probability $\vect{p}^{(t)}$ by DET \eqref{p_det}\;
    Obtain $\tilde{\vect{x}}^{(t)}$ from $\vect{p}^{(t)}$ using \eqref{soft_sym}\;
    Construct $\hat{\vect{H}}_{ob}^{(t)}$ and $\bm{\gamma}^{(t)}$ using \eqref{H_ob} and \eqref{conf}\;
    Stack $\vect{Z}^{(t)}$ according to \eqref{Z_T}\;
    Obtain $\Delta{\mathbf{H}}^{(t)}$ via RefineNet by using \eqref{Delta_H}\;
}

Obtain the final detected data symbols ${\mathbf{\hat x}}_d^{(T)}$ based on a hard decision according to \eqref{x_d_final}\;

\KwOut{Estimated data symbols ${\mathbf{\hat x}_d}$}
\end{algorithm}

The refined CSI is then used for equalization and detection to obtain updated symbol posteriors, which are further utilized to construct the data-aided channel observation for the next iteration, thereby forming an iterative feedback loop. 
The residual formulation restricts RefineNet to learn only a correction to the initial estimate, thereby reducing the learning burden and improving robustness. Meanwhile, the confidence channel provides explicit reliability information across subcarriers, enabling the network to distinguish reliable pilot observations from potentially unreliable data-aided feedback. To further stabilize the optimization, the output layer of RefineNet is initialized to zero such that the refinement starts from the first-pass estimate and progressively improves through iterations.







After the iterative feedback process, a final hard decision is performed based on the resulting symbol posteriors, yielding the detected symbol for each data subcarrier expressed as
\begin{equation}
\hat{x}_{d}[k] = c[{\hat{m}_k}],\label{x_d_final}
\end{equation}
where $\hat{m}_k = \arg\max_{m \in \{1,\ldots,M\}} p_{k,m}^{(T)}$ and 
$T$ denotes the last iteration index. The overall procedure of the proposed CSI RefineNet receiver is summarized in \textbf{Algorithm \ref{alg_refine}}.

\begin{table}[t]
  \centering
  \caption{Module architectures and parameter counts}
  \label{tab_architecture}
  \begin{threeparttable}
  \resizebox{\columnwidth}{!}{%
  \begin{tabular}{llr}
  \toprule
  \textbf{Module} & \textbf{Layer (Configuration)} & \textbf{Params} \\
  \midrule
  \multirow{6}{*}{PilotCE}
   & Conv1D$\times$2: $2{\to}64$, $k{=}3$ + BN + ReLU & 12,864 \\
   & ConvTranspose1D: $64{\to}64$, $k{=}16$, $s{=}8$ & 65,600 \\
   & Conv1D: $64{\to}64$, $k{=}3$ + BN + ReLU & 12,416 \\
   & FC: $64{\to}128$ + Sinusoidal PE & 8,320 \\
   & Transformer Enc.$\times$3 ($d{=}128$, $h{=}4$, $d_{\mathrm{ff}}{=}384$) & 495,360 \\
   & Output FC: $128{\to}2$ & 258 \\
  \cmidrule{2-3}
   & \textbf{Subtotal} & \textbf{595,970} \\
  \midrule
  \multirow{4}{*}{NeuralEQ}
   & Learnable $\beta{=}e^{\theta_\beta}$ & 1 \\
   & Conv1D$\times$2: $6{\to}64{\to}64$, $k{=}5$ + BN + GELU & 22,528 \\
   & Output Conv1D: $64{\to}2$, $k{=}1$ & 130 \\
  \cmidrule{2-3}
   & \textbf{Subtotal} & \textbf{22,915} \\
  \midrule
  \multirow{3}{*}{DET}
   & FC: $2{\to}128$ + Sinusoidal PE & 384 \\
   & Transformer Enc.$\times$2 ($d{=}128$, $h{=}4$, $d_{\mathrm{ff}}{=}256$) & 264,704 \\
   & Output FC: $128{\to}M$ & 516 \\
  \cmidrule{2-3}
   & \textbf{Subtotal} & \textbf{265,860} \\
  \midrule
  \multirow{4}{*}{RefineNet}
   & Conv1D$\times$2: $5{\to}128{\to}128$, $k{=}3$ + BN + ReLU & 51,456 \\
   & FC: $128{\to}128$ + Sinusoidal PE & 16,512 \\
   & Transformer Enc.$\times$3 ($d{=}128$, $h{=}4$, $d_{\mathrm{ff}}{=}384$) & 495,360 \\
   & Output FC: $128{\to}2$ & 258 \\
  \cmidrule{2-3}
   & \textbf{Subtotal} & \textbf{564,738} \\
  \midrule
  \multicolumn{2}{l}{Pass~1 (PilotCE + NeuralEQ + DET)} & \textbf{884,745} \\
  \multicolumn{2}{l}{Pass~2 adds (RefineNet)} & 564,738 \\
  \multicolumn{2}{l}{\textbf{Total (CSI RefineNet receiver)}} & \textbf{1,449,483} \\
  \bottomrule
  \end{tabular}%
  }

  \begin{tablenotes}
    \footnotesize
    \rightskip=2em
    \item \hspace*{-1em} \raisebox{0.2ex}{\scalebox{0.6}{$\bullet$}} $\to$ denotes the mapping from input dimension to output dimension.
    \item \hspace*{-1em} \raisebox{0.2ex}{\scalebox{0.6}{$\bullet$}} BN denotes the batch norm.
    \item \hspace*{-1em} \raisebox{0.2ex}{\scalebox{0.6}{$\bullet$}} ReLU denotes the rectified linear unit activation function.
    \item \hspace*{-1em} \raisebox{0.2ex}{\scalebox{0.6}{$\bullet$}} FC denotes the fully connected layer.
    \item \hspace*{-1em} \raisebox{0.2ex}{\scalebox{0.6}{$\bullet$}} PE denotes the positional encoding.
    \item \hspace*{-1em} \raisebox{0.2ex}{\scalebox{0.6}{$\bullet$}} Transformer Enc. denotes the Transformer encoder.
    \item \hspace*{-1em} \raisebox{0.2ex}{\scalebox{0.6}{$\bullet$}} In the Transformer encoder, $d$ denotes the model dimension, $h$ denotes the number of attention heads, and $d_{\mathrm{ff}}$ denotes the hidden dimension of the feed-forward network.
    \item \hspace*{-1em} \raisebox{0.2ex}{\scalebox{0.6}{$\bullet$}} GELU denotes the Gaussian error linear unit activation function.
    
  \end{tablenotes}
  \end{threeparttable}
\end{table}


\subsection{Module Architectures}
PilotCE takes the real-valued representation of $\vect{y}^T_{\mathcal{P}} \in \mathbb{C}^{1\times |\mathcal{P}|}$ formed by separating its real and imaginary parts, and maps $\mathbb{R}^{2 \times |\mathcal{P}|}$ to $\mathbb{R}^{2 \times N}$. Two Conv1D layers first extract local pilot features, a transposed convolution lifts them to the full set of subcarriers, and a 3-layer Transformer encoder ($d=128$, 4 heads, $d_{\mathrm{ff}}=384$) models long-range frequency-domain structure before a linear layer outputs the real and imaginary channel components. This arrangement lets the network interpolate sparse pilot information while still accounting for global frequency correlation.

NeuralEQ uses a hybrid equalization rule that combines analytical MMSE equalization with a learned residual correction:
\begin{equation}
    \hat{x}_{eq}[k] = \underbrace{\frac{\hat{H}_{eff}^*[k,k] y[k]}{|\hat{H}_{eff}[k,k]|^2 + \beta}}_{\hat{ {x}}^{\mathrm{MMSE}}[k]} + \underbrace{g_{\theta}([\vect{y}, \hat{\vect{H}}_{eff}, \hat{\vect{x}}^{\mathrm{MMSE}}])_k}_{\text{learned ICI correction}},
    \label{eq:neuraleq}
\end{equation}
where $\beta = e^{\theta_\beta}$ is a learnable regularization term and $g_{\theta}$ is a 2-layer convolutional neural network (CNN), where the first layer performs feature expansion ($6 \rightarrow 64$) and the second layer projects the features back to the output space ($64 \rightarrow 2$). The analytical term handles the dominant linear inversion, while the CNN focuses on residual distortion and ICI that remain after model-based equalization.

DET is a 2-layer Transformer encoder ($d=128$, 4 heads) that processes the equalized symbols and outputs class logits in $\mathbb{R}^{N \times M}$. 

RefineNet maps the 5-channel tensor $\vect{Z} \in \mathbb{R}^{5 \times N}$ to a complex residual channel correction through two Conv1D layers ($5 \rightarrow 128$), a 3-layer Transformer encoder, and a zero-initialized output layer. 
The detailed network architectures and parameter counts are listed in {Table~\ref{tab_architecture}}.


\section{Numerical Results}
\label{sec:setup}
We consider a SISO OFDM system with $N=64$ subcarriers and a CP length of $N_{cp}=16$. The carrier frequency is $f_c = 14$ GHz and subcarrier spacing is set to $\Delta f=15$ kHz. The channel follows the 3GPP CDL-C model \cite{3gpp38901} with 24 clusters. The number of channel tap is $L = 4$. 
Unless otherwise stated, the pilot spacing is set to $P_s=4$ and the maximum velocity of the system is $200$ km/h. 


\subsection{Training Procedure}

The optimization is performed in a staged manner rather than trained end-to-end from scratch. Specifically, a two-stage training strategy is adopted, while the proposed receiver performs $T=2$ iterations during inference. In Stage~1, PilotCE, NeuralEQ, and DET are jointly trained with the loss function defined as 
\begin{equation}
    \mathcal{L}_1 = \lambda_H \left\| \hat{\vect{H}}_{{ini}} - \vect{H}_{{eff}} \right\|_F^2 + \mathcal{L}_{\mathrm{CE}}\left(\vect{p}^{(1)}, \vect{y}_{\mathrm{idx}}\right),
    \label{eq:loss1}
\end{equation}
where {$\vect{y}_{\mathrm{idx}}$ contains the ground-truth symbol indices on data subcarriers.} The weight $\lambda_H = 2$ balances channel estimation accuracy and symbol detection, and $\mathcal{L}_{\mathrm{CE}}$ is the cross-entropy loss on data subcarriers. This stage establishes a stable pilot-driven detector before any decision feedback is introduced.

In Stage~2, the full two-pass receiver with $T=2$ iterations is unrolled and jointly optimized by
\begin{equation}
    \begin{aligned}
        \mathcal{L}_2 ={}& \left\| \hat{\vect{H}}_{{ini}} - \vect{H}_{eff} \right\|_F^2
        + 3 \left\| \hat{\vect{H}}_{{eff}} - \vect{H}_{eff} \right\|_F^2 \\
        &+ 0.3 \mathcal{L}_{\mathrm{CE}}\left(\vect{p}^{(1)}, \vect{y}_{{idx}}\right) + \mathcal{L}_{\mathrm{CE}}\left(\vect{p}^{(2)}, \vect{y}_{{idx}}\right).
    \end{aligned}
    \label{eq:loss2}
\end{equation}
The larger weight assigned to the refined channel term encourages the second pass to explicitly improve the CSI accuracy, while the reduced weight on the first-pass detection loss prevents the feedback path from dominating the optimization. In practice, this schedule matters because training RefineNet on poorly calibrated early decisions tends to destabilize convergence. Both stages use AdamW with cosine annealing and gradient clipping at norm 1.0, and training samples are drawn from uniformly mixed signal-to-noise ratio (SNR) values of $\{0, 5, 10, 15, 20, 25\}$~dB.

Models are trained separately for each velocity so that the comparison isolates receiver behavior at a fixed mobility level rather than averaging across heterogeneous Doppler conditions. Each run uses 20{,}000 training symbols and 2{,}000 validation symbols, with evaluation on 30{,}000 test symbols. Stage~1 is trained for $E_1=50$ epochs at learning rate $2\times10^{-4}$, and Stage~2 for $E_2=60$ epochs at $10^{-4}$. All neural models use batch size 512 and AdamW with weight decay $3\times10^{-4}$.

\subsection{Simulation results for BER performance}
For comparison, we consider the conventional pilot-aided receivers LS+ZF \cite{edfors1998ofdm} and MMSE+ZF \cite{edfors1998ofdm}, where LS and MMSE are used for channel estimation, respectively, followed by zero-forcing (ZF) equalization for symbol detection.
We also include MMSE+GAMP \cite{GAMP_8746319}, which performs iterative Bayesian detection initialized by the MMSE channel estimation for 15 iterations, and an end-to-end neural baseline DeepRx \cite{honkala2021deeprx}, implemented as a ResNet with 256 channels, three dilated residual blocks, and a 5-channel input formed from the received signal and pilot mask. In addition, a Perfect-CSI setting is included by feeding the true channel response to NeuralEQ and DET.


\begin{figure}
    \centering
    \includegraphics[width=0.9\linewidth]{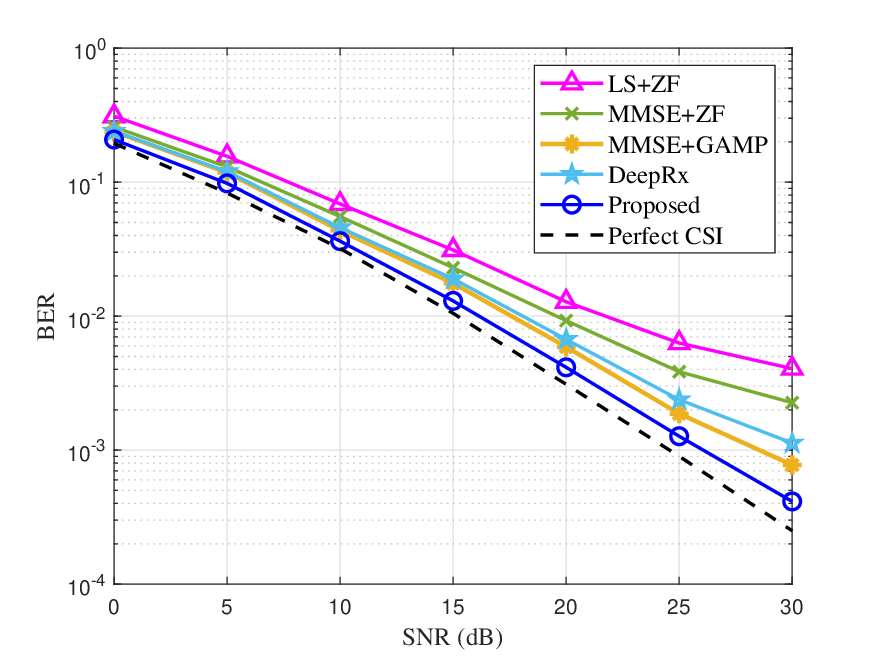}
    \caption{QPSK BER versus SNR with the maximum velocity of 200~km/h and pilot spacing $P_s=4$.}
    \label{BER_200}
\end{figure}

Fig.~\ref{BER_200} depicts the BER versus SNR for different receivers. As the SNR increases, the BER of all methods decreases monotonically. LS+ZF \cite{edfors1998ofdm} performs the worst due to the sensitivity of LS-based pilot CSI to noise and interpolation errors, together with the inability of diagonal ZF detection to suppress Doppler-induced ICI. MMSE+ZF \cite{edfors1998ofdm} improves the BER through more accurate CSI estimation, but remains limited by the same diagonal equalization structure. DeepRx \cite{honkala2021deeprx} benefits from data-driven detection, while MMSE+GAMP \cite{GAMP_8746319} achieves further improvement through iterative Bayesian inference that better exploits the signal structure and partially mitigates residual interference. However, neither method explicitly refines the intermediate CSI, and both therefore remain limited in handling the mismatch between the pilot-based CSI and the effective channel under severe Doppler-induced ICI. In contrast, our proposed CSI RefineNet receiver consistently achieves the lowest BER among all practical receivers and remains the closest to the perfect CSI, owing to its stronger ability to explicitly reduce CSI mismatch and compensate for channel distortion beyond the diagonal approximation.

\begin{figure}
    \centering
    \includegraphics[width=0.9\linewidth]{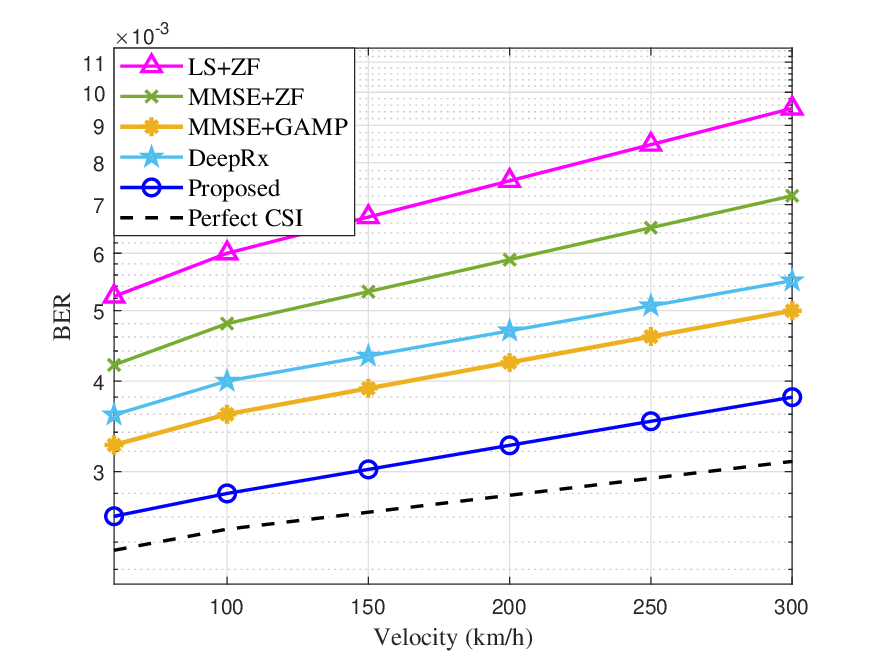}
    \caption{QPSK BER versus velocity at $20$~dB with pilot spacing $P_s=4$.}
    \label{BER vs velocity}
\end{figure}
\begin{figure}
    \centering
    \includegraphics[width=0.9\linewidth]{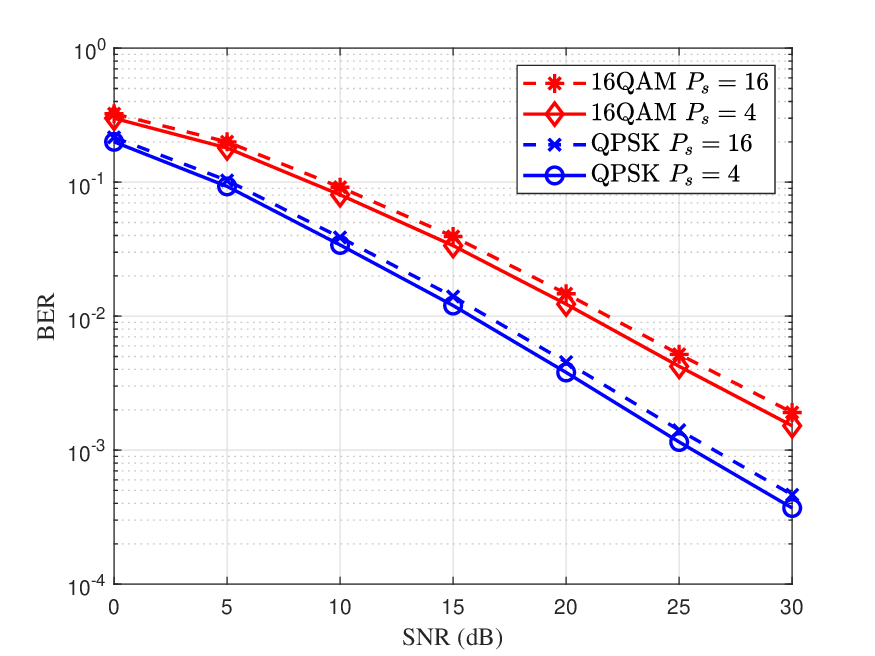}
    \caption{BER versus SNR under different modulation orders and pilot spacings.}
    \label{16QAM}
\end{figure}

Fig.~\ref{BER vs velocity} illustrates the QPSK BER versus velocity at $20$ dB with pilot spacing $P_s=4$. 
The BER of all receivers increases with velocity due to the stronger channel variation and Doppler-induced ICI in high-mobility scenarios. However, our proposed CSI RefineNet receiver consistently achieves the lowest BER across the entire velocity range. This is because the proposed iterative refinement receiver can better reduce the increasing mismatch between the pilot-based CSI and the effective channel on data subcarriers, thereby maintaining more reliable detection performance as mobility grows.


Fig.~\ref{16QAM} shows the BER performance of the proposed CSI RefineNet receiver under different modulation orders and pilot spacings. Under the same pilot spacing, QPSK consistently outperforms 16 QAM, because the denser constellation of 16 QAM leads to a smaller decision margin and hence stronger sensitivity to noise, residual CSI mismatch, and Doppler-induced interference. In addition, $P_s=4$ achieves slightly lower BER than $P_s=16$ for both modulation schemes, since denser pilots provide more accurate initial CSI. The relatively small gap between the two pilot spacing scenarios further indicates that the proposed receiver can effectively alleviate the CSI degradation caused by sparse pilots through iterative data-aided CSI refinement.


\vspace{-0.02in}
\subsection{Complexity Analysis}
\vspace{-0.12in}
\begin{table}[h]
\centering
\caption{Complexity and memory comparison.}
\label{complexity_memory}
\begin{tabular}{lcc}
\hline
\textbf{Method} & \textbf{FLOPs} & \textbf{Total Memory} \\
\hline
LS + ZF            & 166.4K & --     \\
MMSE + ZF          & 4.9M   & --     \\
MMSE + GAMP        & 40.3M  & --     \\
DeepRx             & 37.8M  & 5.0 MB \\
Proposed           & 34.1M  & 4.2 MB \\
\hline
\end{tabular}
\end{table}

Table \ref{complexity_memory} compares the computational complexity and memory consumption of the considered receivers. LS+ZF and MMSE+ZF, as two simple linear receivers, have relatively low complexity but suffer from worse BER performance compared with that of the other schemes. MMSE+GAMP receiver achieves better BER performance than these linear baselines, but the improvement comes at the cost of a significantly increased complexity of 40.3M FLOPs. DeepRx and the proposed scheme, as two learning-based receivers, both introduce additional computational and memory overhead. Compared with DeepRx, the proposed scheme achieves better BER performance while reducing the computational complexity from 37.8M to 34.1M FLOPs and the memory consumption from 5.0 MB to 4.2 MB. Such result demonstrates that better receiver performance usually comes with increased computational and memory cost, while the proposed method achieves a better tradeoff among these factors.

\vspace{0.5em}
\section{Conclusion}
\label{sec:conclusion}
In this paper, we proposed a CSI RefineNet receiver for high-mobility OFDM systems to address the mismatch between pilot-based channel state information and the effective channel experienced by data subcarriers in rapidly time-varying channels. The proposed receiver started from a pilot-driven coarse CSI estimation, and then exploited soft symbol posteriors to construct data-aided channel observations for iterative CSI refinement through a dedicated residual refinement network. By feeding the refined CSI back to the equalization and detection modules, a closed-loop receiver structure was established. In addition, a two-stage training strategy was developed to improve optimization stability and fully exploit the benefit of iterative refinement. Simulation results showed that the proposed method consistently achieved the best BER performance among the considered benchmark receivers and remained robust under different velocities, modulation orders, and pilot-spacing configurations. 
\section*{Acknowledgment}
This work was supported by A*STAR under the RIE2025 Industry Alignment 
Fund–Industry Collaboration Projects (IAF-ICP) Funding Initiative (Award: I2501E0045), as well as cash and in-kind contribution from the industry partner(s).


\bibliographystyle{IEEEtran}
\footnotesize
\bibliography{reference}

\end{document}